\documentclass[preprint]{aastex63}
\usepackage{CJK}
\received{----}
\revised{----}
\accepted{February 15, 2022}
\submitjournal{ApJ}

\shorttitle{Cepheids' pulsation mode v.s. metallicity}
\shortauthors{Zhang et al.}

\begin{document}
\begin{CJK*}{UTF8}{gbsn}
\title{Dependence of pulsation mode of Cepheids on metallicity}

\correspondingauthor{Biwei Jiang}
\email{bjiang@bnu.edu.cn}

\author[0000-0002-3828-9183]{Zehao Zhang (张泽浩)}
\affiliation{Department of Astronomy, Beijing Normal University, Beijing 100875, China}

\author[0000-0003-3168-2617]{Biwei Jiang (姜碧沩)}
\affiliation{Department of Astronomy, Beijing Normal University, Beijing 100875, China}

\author[0000-0003-1218-8699]{Yi Ren (任逸)}
\affiliation{College of Physics and Electronic Engineering, Qilu Normal University, Jinan 250200, China}

\author[0000-0001-7084-0484]{Xiaodian Chen (陈孝钿)}
\affiliation{CAS Key Laboratory of Optical Astronomy, National Astronomical Observatories, Chinese Academy of Sciences, Beijing 100101, China}
\affiliation{Department of Astronomy, China West Normal University, Nanchong 637009, China}
\affiliation{School of Astronomy and Space Science, University of the Chinese Academy of Sciences, Huairou 101408, China}

\author[0000-0003-4489-9794]{Shu Wang (王舒)}
\affiliation{CAS Key Laboratory of Optical Astronomy, National Astronomical Observatories, Chinese Academy of Sciences, Beijing 100101, China}
\affiliation{Department of Astronomy, China West Normal University, Nanchong 637009, China}

\begin{abstract}
The Cepheid variables in SMC, LMC, the Milky Way, M33 and M31 are used to examine the dependence of pulsation mode on metallicity which was previously found in red supergiants. The initial samples of Cepheids are collected from the Cepheid catalogs identified from the OGLE, PS1, DIRECT, WISE and ZTF surveys. The contaminants are removed with the help of the Gaia/EDR3 astrometric information for extra galaxies or by comparing the geometric distance and the distance from the P-L relation for the Milky Way. The division of fundamental and first-overtone mode is refined according to the gap between the two modes in the P-L diagram of the objects in each galaxy. The ratio of FU/(FU+1O) is found to be 0.59, 0.60, 0.69, 0.83 and 0.85 for SMC, LMC, the Milky Way, M33 and M31 respectively in order of metallicity, which confirms that the pulsation mode depends on metallicity in the way that the ratio of FU/(FU+1O) increases with metallicity. This dependence is not changed if the incompleteness of the samples is taken into account.

\end{abstract}

\keywords{Cepheid variable stars (218); Pulastion modes (1309); Metallicity (1031)}


\section{Introduction} \label{sec:1}

In analyzing the light variation of red supergiants (RSGs) in M31 and M33, \citet{2019ApJS..241...35R} identified the pulsation mode by matching the observationally determined period-luminosity relation (PLR) with the theoretical calculations. When combing with the results of SMC, LMC and the Milky Way, they found that the number ratio of fundamental mode (FU) to first-overtone (1O) depends on metallicity in the way that it increases with metallicity, i.e. there are more RSGs pulsating in the FU mode than in the 1O mode at high metallicity. They explained this dependence of pulsation mode of RSGs by the metallicity effect on mixing length. \citet{2018ApJ...853...79C} suggested that the mixing length increases with the metallicity for RSGs. Meanwhile for Cepheids, \citet{1998A&A...336..553Y} found that the increase of the mixing length would greatly inhibit the relative growth rate of the 1O, thus reducing the instability strip of the 1O more sharply. Because convection is popular in RSGs (see e.g. \citet{2020ApJ...898...24R}), the damping of 1O mode is expected as well. The combined effect of the two factors would result in a rising fraction of the FU mode pulsators at high metallicity.

The dependence of pulsation mode on metallicity needs further confirmation. It is well known that the light variation of RSGs is generally not regular, which brings about large uncertainty in the derived period. Moreover, the sample of variable RSGs in the study of \citet{2019ApJS..241...35R} is far from complete \citep{2021ApJ...907...18R}. The additional uncertainty arises from the theoretical calculation such as the selection of the mixing length \citep{2002ApJ...565..559G}.

This work chooses classical Cepheids to examine the dependence of pulsation mode on metallicity. Cepheids are mostly regular variables with well defined PLR. Supplemented by high luminosity, Cepheids are easily recognized even outside the Local Group Galaxies \citep{1988ApJ...332L..63F, 2011ApJ...743..176G} and  used to determine the Hubble constant \citep{2001ApJ...553...47F, 2018ApJ...861..126R, 2019ApJ...876...85R}. Cepheids pulsate in radial mode, including the fundamental, first-overtone, second-overtone mode, and some pulsate in double mode \citep{2013ApJ...778..147P, 2015AcA....65..329S, 2018AcA....68..341J}, but the majority pulsate in the fundamental and first-overtone mode. Thus Cepheids are a much better type of variable to examine the dependence of pulsation mode on metallicity than RSGs.

This work will analyze the FU/(FU+1O) ratio with the Cepheid sample in the Magellanic Clouds, the Milky Way, M33, and M31. Indicated by the oxygen abundance $12+\log(O/H)$, the metallicity of these galaxies covers a wide range from sub-solar in SMC and LMC to solar in the Milky Way and M33, and then to super-solar in M31.

\section{The Data} \label{sec:Data}

The numbers of Cepheids pulsating in FU and 1O mode respectively are needed to study the FU/(FU+1O) ratio. This is carried out in different ways depending on the galaxy individually.

\subsection{SMC and LMC}

The Magellanic Clouds are the nearest neighbor of the Milky Way with a lot of observations. The time-domain observation by the Optical Gravitational Lensing Experiments (OGLE; \citet{2015AcA....65....1U}) is an excellent project to study the variable stars. From stellar light variation collected for a long duration at an appropriate cadence by OGLE, \citet{2015AcA....65..297S} identified 4,915 and 4,620 Cepheids in SMC and LMC, respectively. According to their identification, these Cepheids in SMC include 2,754 and 1,791 objects pulsating in the FU and 1O mode respectively, and the Cepheids in LMC include 2,477 and 1,776 objects pulsating in the FU and 1O mode respectively.

\subsection{M33 and M31}

The sample of Cepheids in M33 is taken from \citet{2011ApJS..193...26P}. They combined the data from the DIRECT project that is a dedicated survey of eclipsing binaries and Cepheids in M31 and M33 by the observations collected at the Michigan-Dartmouth-MIT Observatory 1.3-m telescope and the F. L. Whipple Observatory 1.2-m telescope \citep{1998AJ....115.1894S}, and WIYN \citep{2011ApJS..193...26P}. \citet{2011ApJS..193...26P} built a sample of 698 Cepheids in M33, 563 of which are identified as classical Cepheids.

The sample of Cepheids in M31 is taken from \citet{2018AJ....156..130K}. Although the DIRECT project also observed M31 along with M33, \citet{2011ApJS..193...26P} found only several tens of Cepheids in M31. Instead, \citet{2018AJ....156..130K} presented the largest Cepheid sample in M31 based on the complete Pan-STARRS1 survey \citep{2016arXiv161205560C} of Andromeda  in the $r_{\rm P1}$, $i_{\rm P1}$, and $g_{\rm P1}$ bands. They detected 2,686 Cepheids, among which 1,662 pulsates in the FU mode, 307 in the 1O mode, and some belong to other types.


\subsection{The Milky Way}

The Milky Way differs from the external galaxies in that there is no possibility to achieve a complete spatial coverage due to our location. Consequently, the sample of Cepheids can only be collected in some sky regions. Nevertheless, recent observations and studies have obtained a few typical samples of Cepheids. Our data of the Galactic Cepheids incorporates the samples from the WISE, ZTF, and OGLE surveys.

Based on the all-sky survey by the Wide-field Infrared Survey Explorer (WISE; \citet{2010AJ....140.1868W}) in the W1 (3.35 $\mu$m), W2 (4.60 $\mu$m), W3 (11.56 $\mu$m) and W4 (22.09 $\mu$m) band, \citet{2018ApJS..237...28C} identified 867 classical Cepheids in the Milky Way by its five-year survey data. Since the amplitude of the 1O is about half of the FU, difficult to detect in the WISE mid-infrared bands, these Cepheids cannot be of the 1O mode \citep{2019NatAs...3..320C}. In addition, according to our preprocessing of this part of data,  not all the 867 Cepheids are the FU mode Cepheids, and further analysis will be presented in Section \ref{sec:3.2}.

From the Zwicky Transient Facility (ZTF; \citet{2019PASP..131f8003B, 2019PASP..131a8003M}) time-domain survey of almost the entire northern sky that began in 2018 with a 47-square-degree field of view camera mounted on the 48-inch Schmidt telescope in Palomar, \citet{2020ApJS..249...18C} identified 1,262 classical Cepheids in the Milky Way using the ZTF/DR2 database, of which 565 are newly discovered.

In addition, \citet{2018AcA....68..315U} presented the OGLE collection of 2,721 Cepheids of all types in the Galactic disk and outer bulge, among which there are 1,201 FU mode and 551 1O mode Cepheids.

\section{Refinement of the classification of the pulsation mode}

\subsection{Removal of the foreground stars for SMC, LMC, M33 and M31}

\begin{figure}[htb]
    \plotone{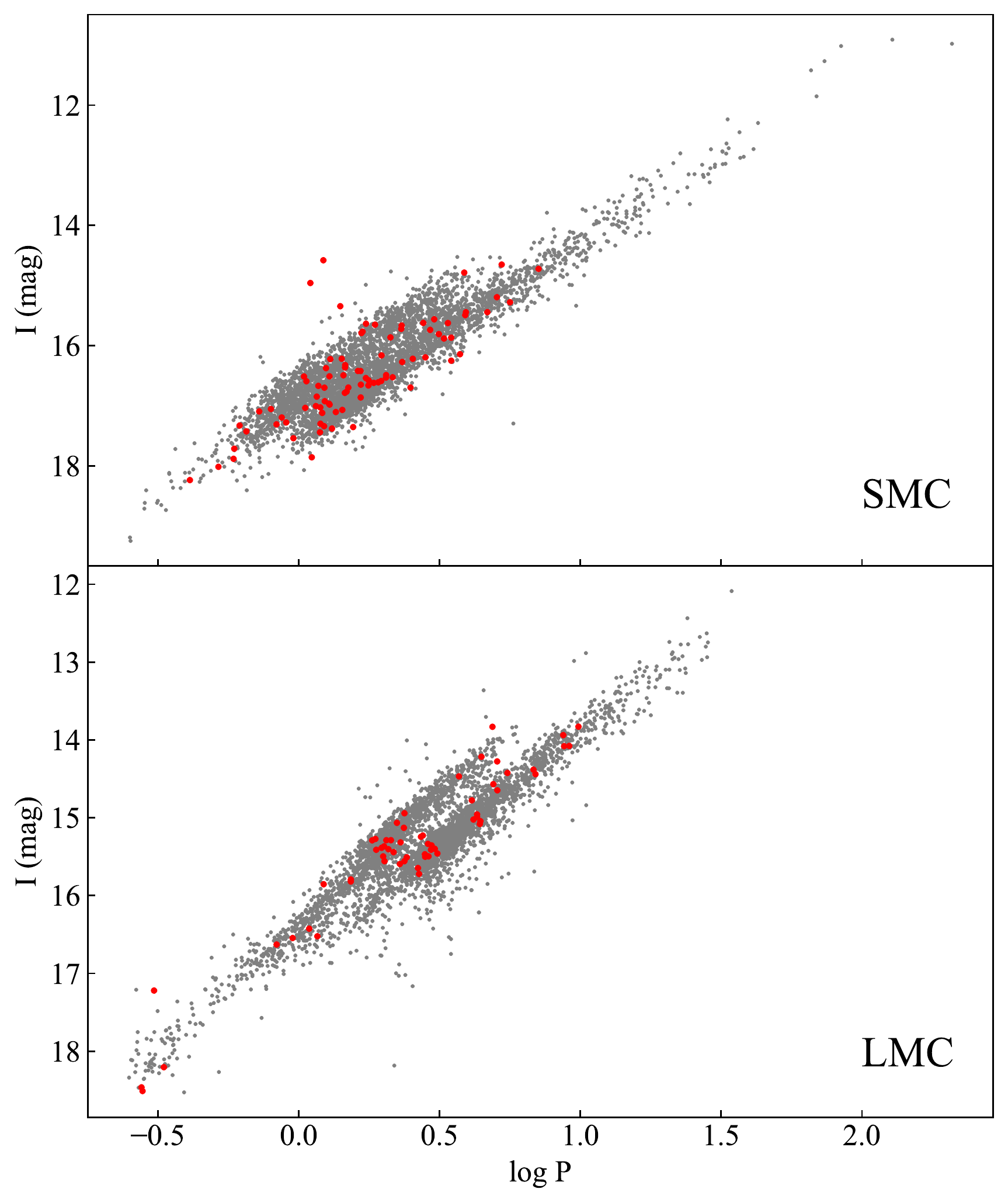}
    \caption{The P-L diagram of Cepheids in SMC (upper) and LMC (lower), where the red dots denote the foreground stars judged from the Gaia measurements of the parallax or proper motion.  \label{fig:sl_fgd}}
\end{figure}

The stars in the samples are examined with the Gaia astrometric information preliminarily in order to confirm their membership. The criteria is the same as that introduced in \citet{2019A&A...629A..91Y}, \citet{2020A&A...639A.116Y}, and \citet{2021arXiv211008793R}. In brief, a star in SMC or LMC is considered to be a foreground star if its proper motion is beyond $5\sigma$ of the peak value of the host galaxy or its distance is closer than 30 kpc. The astrometric data is retrieved from Gaia/EDR3 \citep{2021A&A...649A...1G}, in which only the measurements with relative error less than 20\% are taken into account. In this way, 87 and 60 stars in the SMC and LMC sample respectively are identified as foreground stars. However, their locations in the P-L diagram, as shown by red dots in Figure \ref{fig:sl_fgd}, coincide very well with the other member stars, except that two stars in SMC are brighter by about $1\sim$mag than the stars at the same period. Since a foreground Cepheid would appear much brighter than the member one and deviate significantly from the PLR, and because of the large thermal oscillation Gaia received after launch, we can't guarantee that the astrometric data for each star is accurate \citep{2021ApJ...908L...6R}. Despite their high confidence in statistical significance (e.g., 99\% of the stars in the catalog are measured reliably), we decide not to remove them and speculate that they might be Cepheid binaries or Cepheids with abnormal metallicity \citep{2012A&A...537A..81S}. For the sample stars in M33 and M31, the Gaia/EDR3 astrometric data are available for 334 and 857 stars respectively, and none of them show a proper motion or parallax to be like a foreground star. This examination confirms the membership of the sample stars.

\subsection{Division of the fundamental and first-overtone mode}

\begin{figure*}[htb]
    \plotone{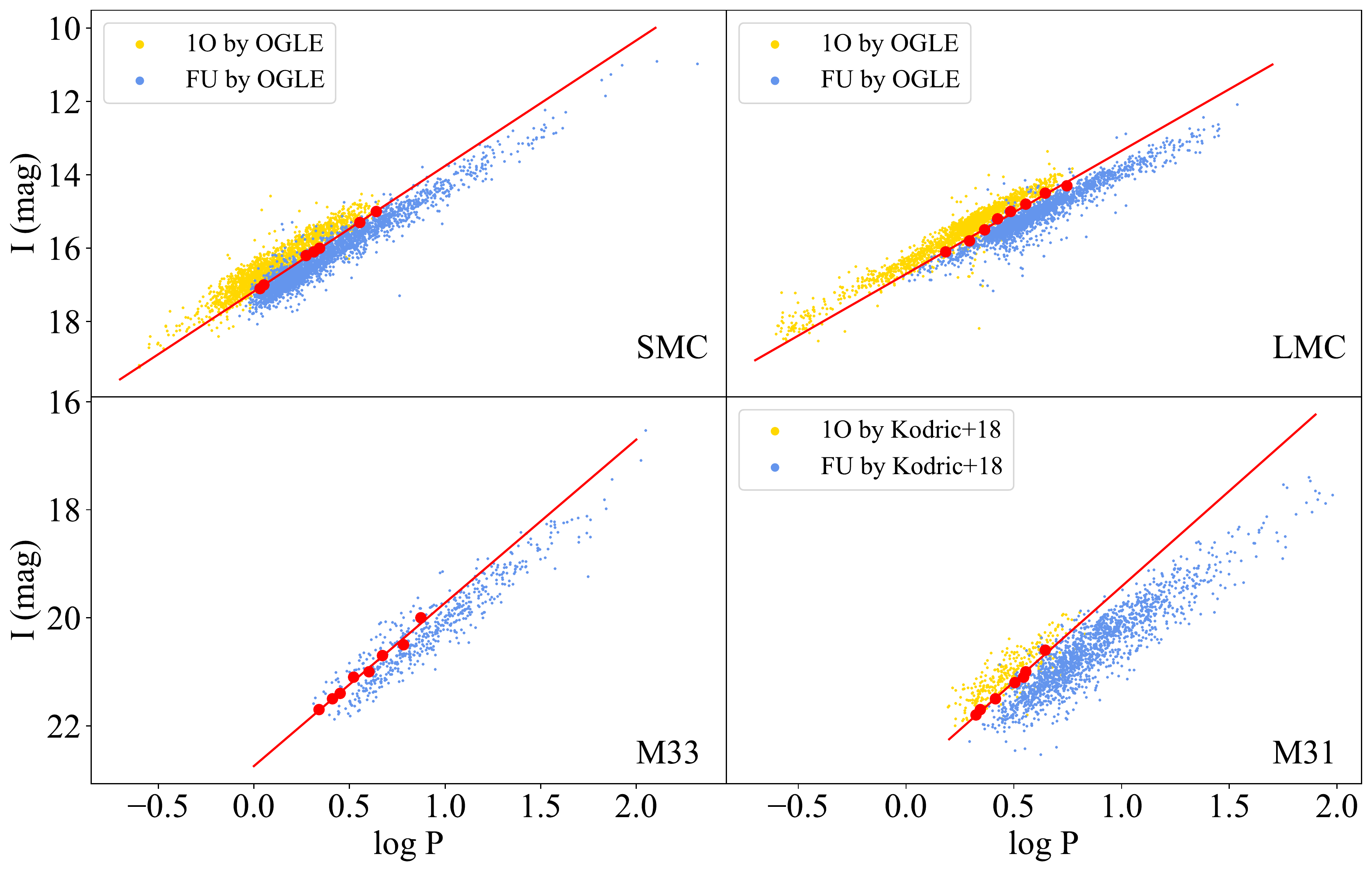}
    \caption{Refinement of the pulsation mode by the location in the $\log P$ - I diagram, with the red dots and red line being the troughs at selected magnitudes and division line between the FU and 1O mode, respectively. The yellow and blue dots denote the 1O and FU mode respectively identified in previous references for SMC, LMC \citep{2015AcA....65..297S} and M31 \citep{2018AJ....156..130K}. The pulsation mode of Cepheids in M33 is newly defined in this work.  \label{fig:division}}
\end{figure*}

The previous pulsation mode identification of the Cepheids in the sample were based on the amplitude ratio and phase difference yielding from Fourier analysis in the sub-components of the light curves. This may lead to some uncertainty for noisy light curves of distant stars, in particular for the objects in M31 and M33. The misidentification of the mode is evidenced by the outliers in the P-L diagram, as illustrated in Figure \ref{fig:division} where some FU stars locate within the P-L ranges of 1O stars and vice versa. Because the period and the magnitude are usually more accurately measured than the amplitude ratio and phase difference, we decide to refine the identification of the pulsation mode by the location in the P-L diagram. As Cepheids pulsating in FU and 1O mode show two distinct branches in the P-L diagram, a division line is determined to distinguish between the two pulsation modes. Firstly, the magnitude is sliced with an appropriate bin, which yields a two-peak histogram along the period in a specific bin that correspond to the most densely distributed $\log P$ values of the FU and 1O mode respectively. Then a double-Gaussian fitting on the distribution of the numbers is performed, and the resultant position of the trough between the two Gaussians is taken as the dividing position between the FU and 1O mode at this magnitude. After repeating such operation for a number of selected magnitudes showing significant gap between the two modes, a linear fit to these points determines the division line that distinguishes between the FU and 1O mode. The division line with the sample stars are shown in the P-L diagram (Figure \ref{fig:division}). There are 4 and 37 sample stars in SMC and LMC respectively for which the I-band photometric data are lack, consequently our method is invalid to identify their pulsation mode. In such case, we accept the conclusion of \citet{2015AcA....65..297S} that consider them all being the FU pulsator. It is worth noting that the photometry of the I-band is used for the four galaxies although the data was taken by different facilities because the I-band interstellar extinction is about half of that in the V-band \citep{2019ApJ...877..116W}. In addition, the V-band photometric points are fewer than the I-band. For example, 185 and 232 Cepheids in SMC and LMC have no V-band photometric data, respectively. The number of Cepheids pulsating in FU and 1O mode in four galaxies is shown in Table \ref{tab:final}. In this way, 87 stars in SMC, 70 stars in LMC and 15 stars in M31 are identified with different pulsation mode from previous results, meanwhile the classification is newly done for the objects in M33.


\subsection{Integration of the Galactic sample}\label{sec:3.2}

\begin{figure*}[htb]
    \gridline{\fig{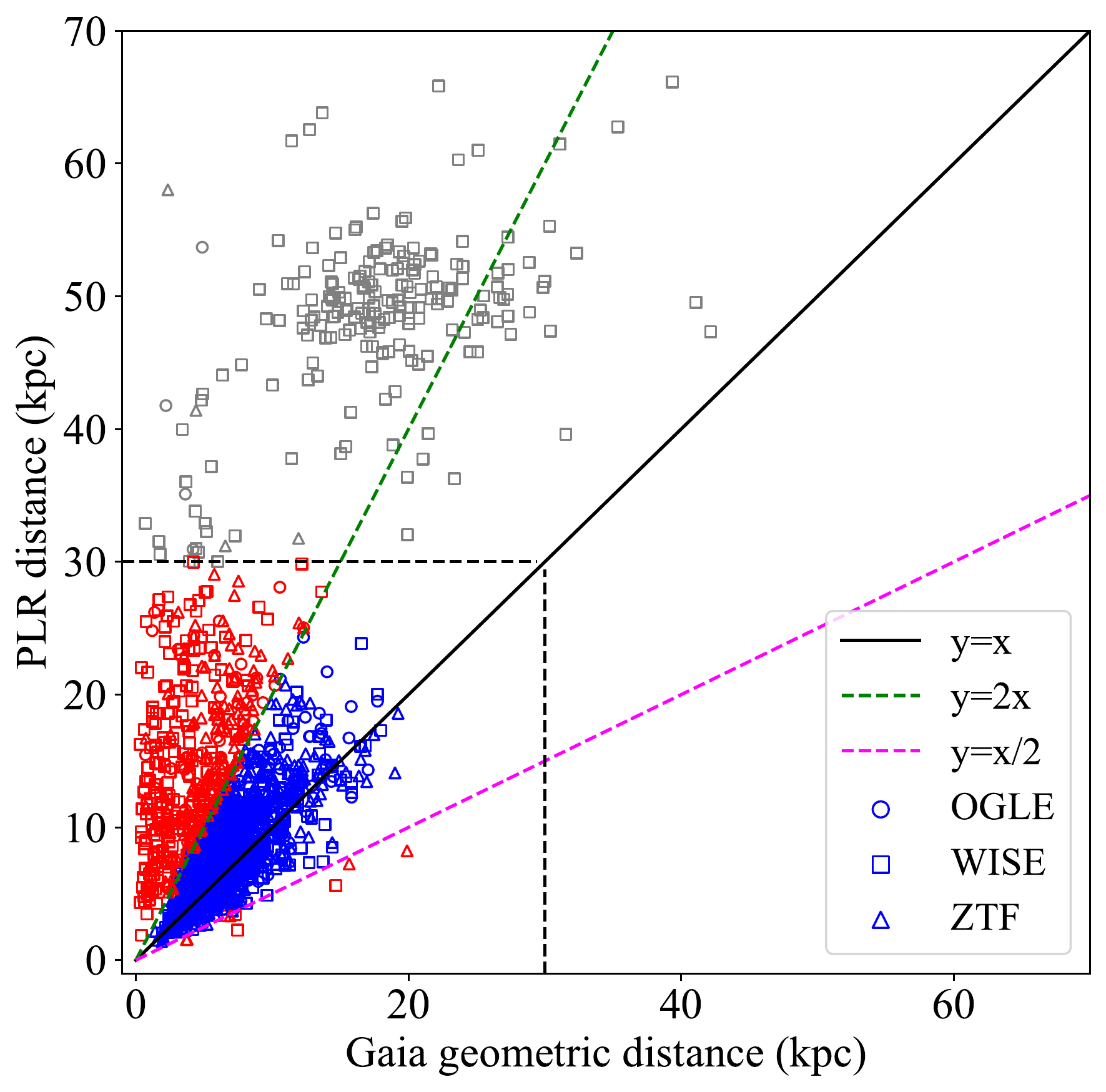}{0.45\textwidth}{(a)}
              \fig{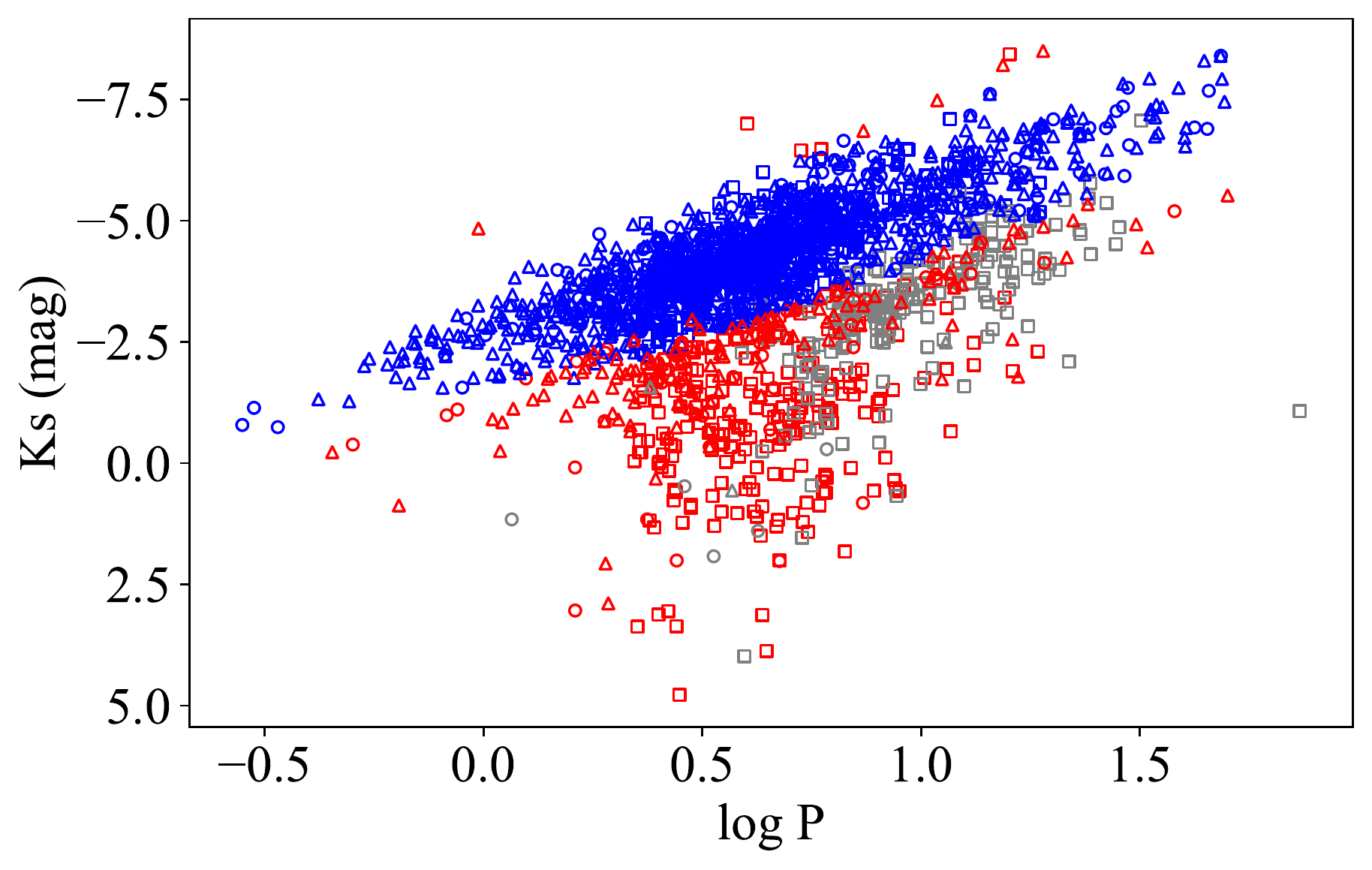}{0.56\textwidth}{(b)}
              }
    \gridline{\fig{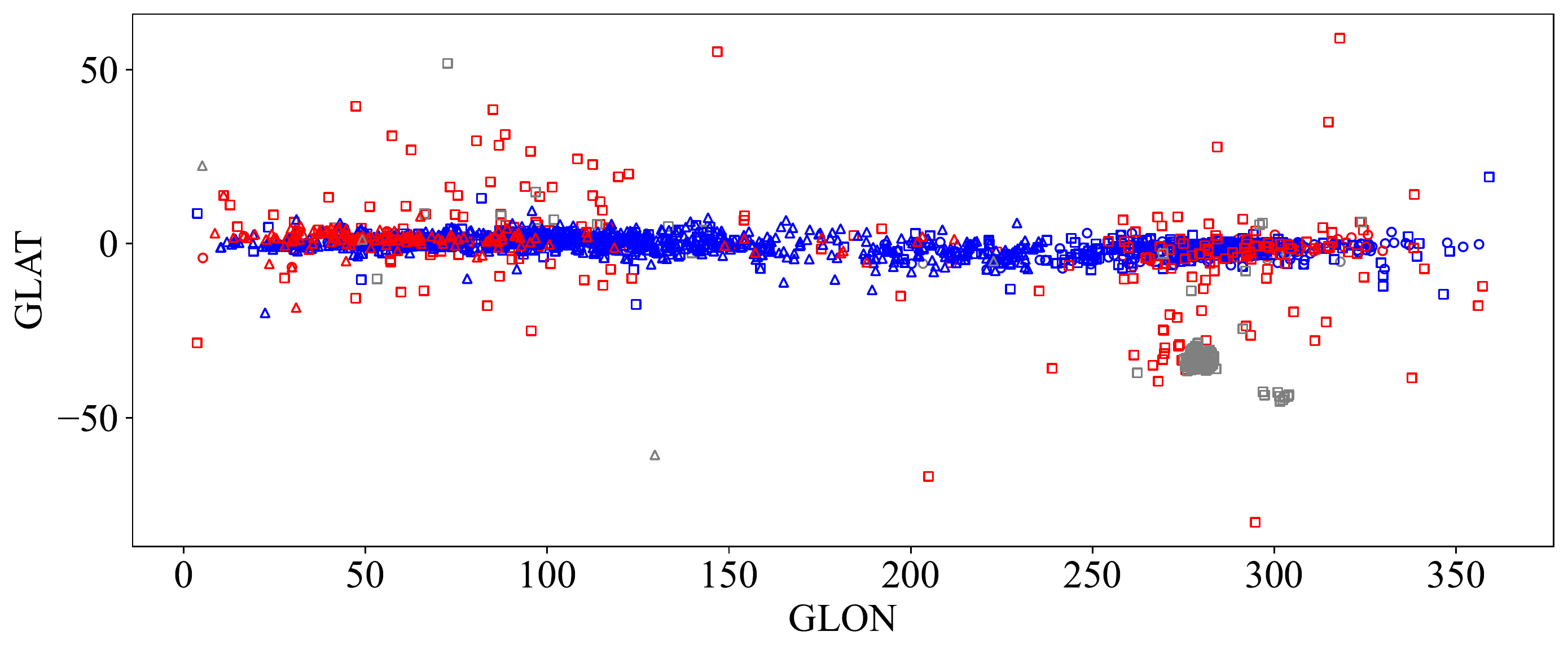}{0.8\textwidth}{(c)}
              }
    \caption{Purification of the Cepheids from the WISE (square) \citep{2018ApJS..237...28C}, ZTF (triangle) \citep{2020ApJS..249...18C} and OGLE (circle) \citep{2018AcA....68..315U} catalog. The blue symbols denote the accepted,  the red denote the excluded stars, and the grey denote the stars with the PLR or Gaia geometric distances greater than 30 kpc. Figure (a) compares the PLR distance and the geometric distance from Gaia by \citet{2021AJ....161..147B}, Figure (b) and (c) display their $\log P -$ K-mag diagram and distribution in the Galaxy respectively. \label{fig:milky_way}}
\end{figure*}

Due to our location inside the Galactic plane with severe extinction, there can be some contaminants in the catalogs of Galactic Cepheids, mainly type-II Cepheids and rotating stars, and even some extragalactic sources. The identifications are further examined by comparing the distance from the $K_{\rm S}$-band P-L relation of Cepheids (the PLR distance) with the geometric distance from \citet{2021AJ....161..147B}. Geometric distance is the stellar distance calculated by \citet{2021AJ....161..147B} that was based on a prior 3-D model of the Milky Way including Gaia's apparent magnitude, color, parallax, etc., for 1.47 billion stars by using the Bayesian method. First the PLR distance is calculated from the apparent magnitude ($m_{K_{\rm S}}$) and the absolute magnitude in the $K_{\rm S}$-band ($M_{K_{\rm S}}$) with the PLR of \citet{2017MNRAS.464.1119C} from the relation ${5}\lg_{}{r} =m_{K_{\rm S}}-M_{K_{\rm S}}+5-A_{K_{\rm S}}$. For Cepheids pulsating in FU mode, we directly adopt the PLR calculated by \citet{2017MNRAS.464.1119C}. For Cepheids pulsating in 1O mode, since they are about 0.5 magnitude brighter than the stars in FU mode \citep{2020ApJS..249...18C}, the value of $M_{K_{\rm S}}$ is shifted downward by 0.5 mag. The $A_{K_{\rm S}}$ is converted from the color excess $E(H-W2)$ with the extinction law of \citet{2019ApJ...877..116W}, i.e. $A_{K_{\rm S}}= 0.73 E(H-W2)$. Moreover, the color excess $E(H-W2)$ is calculated by assuming that the intrinsic color index $(H-W2)_0$ is a constant of 0.08 for Cepheids mainly of spectral type F6-K2 as demonstrated by \citet{2011ApJ...739...25M}. The exclusion criterion is the same as in \citet{2020A&A...639A..72W}, i.e. a star is considered to be a Galactic Cepheid when it lies between the lines $y=x/2$ and $y=2x$ in the PLR distance v.s. Gaia geometric distance diagram shown in Figure \ref{fig:milky_way} and its distance is smaller than 30 kpc.

This method is applied to the Cepheid catalogs from the WISE, ZTF, and OGLE surveys. The WISE catalog is an apparently contaminated sample, with 444 stars removed, leaving 423 classical Cepheids. For the ZTF and OGLE catalogs, 1,144 and 332 stars are successfully cross-matched respectively with the Gaia geometric distance and the $H$, $K_{\rm S}$, $W2$ band photometry, among which 149 and 59 stars are removed respectively. It can be seen in Figure \ref{fig:milky_way} that the removed stars deviate apparently from the $K_{\rm S}$-band PLR that they should follow, and they are mostly located at high Galactic latitude. Furthermore, some of the removed stars from the WISE catalog appear in the SMC and LMC sky area. Indeed, the cross-match with the OGLE catalog of Cepheids in SMC and LMC finds 8 and 138 objects respectively, meanwhile the WISE stars confirmed to be classical Cepheids are not detected in the SMC or LMC catalogs, which supports the correctness of our criteria.

As for the mode analysis, of the 193 Cepheids present in both the OGLE and ZTF catalogs after excluding the contaminants, 4 are classified into different pulsation modes. Despite the contradiction, we still adopt the identification results of ZTF. Based on the surveys of these two facilities, there are 1,789 and 824 Cepheids pulsating in FU and 1O mode, respectively. Three out of the 205 Cepheids present in both the WISE and OGLE \& ZTF catalogs are classified as 1O mode by the latter, but we still assume them to be in the FU mode as the 1O pulsation would be very difficult to detect in the WISE infrared band. In addition, the WISE catalog is certainly biased to the FU mode, which would lead to over-estimating the ratio FU/(FU+1O) for the Milky Way. So this catalog is used only for confirming and correcting the mode classification while it is not included in the following analysis. After excluding the duplicated sources, the final number of Cepheids in the FU and 1O modes in the Milky Way are 1,792 and 821 respectively as listed in Table \ref{tab:final}.

\section{Result and discussion}\label{sec:result}

\subsection{Dependence of the number ratio FU/(FU+1O) on metallicity}

\begin{figure}[htb]
    \plotone{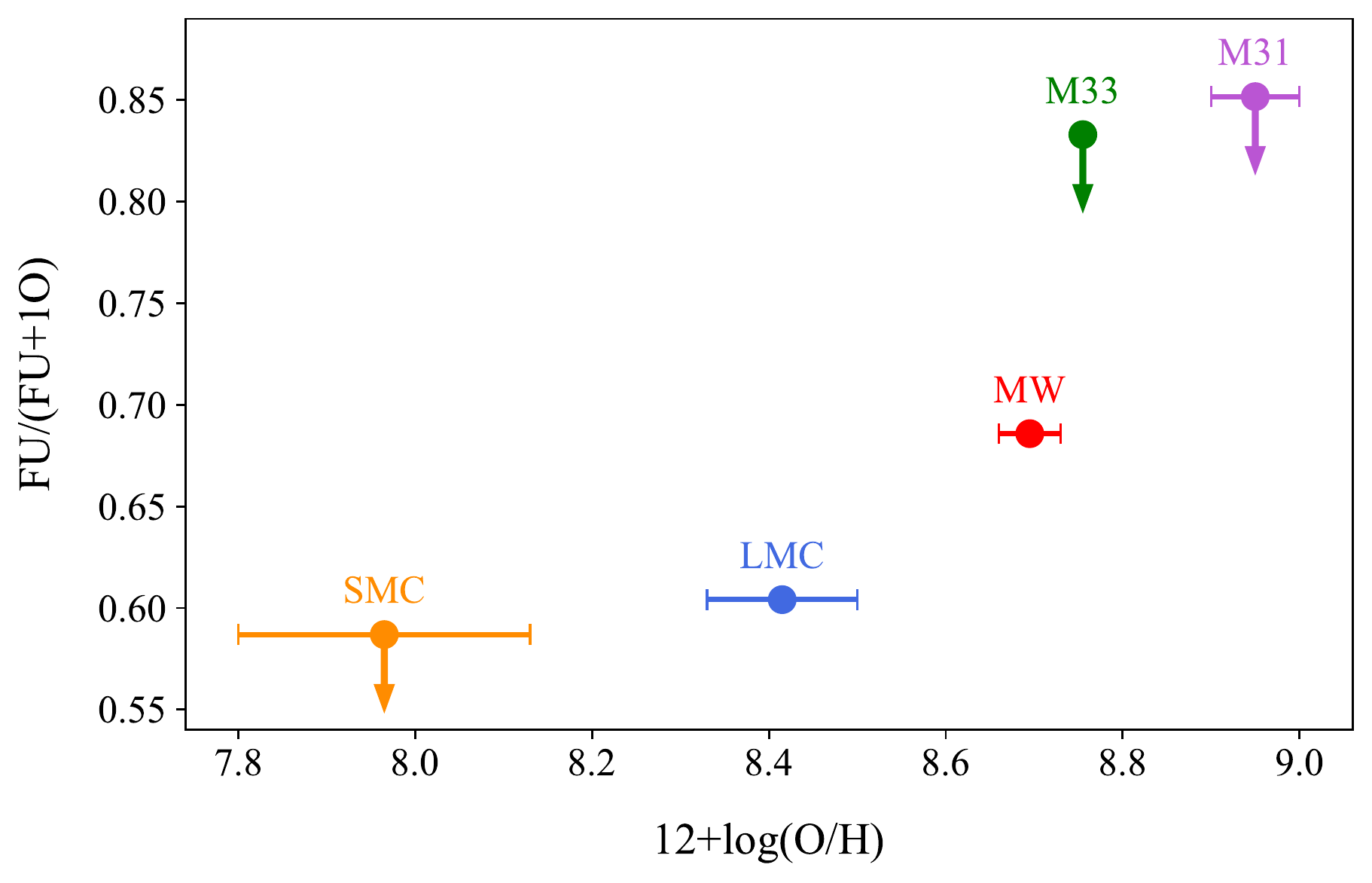}
    \caption{The FU/(FU+1O) ratio of Cepheids versus the metallicity represented by 12+$\log$(O/H) in the target galaxies. The arrows indicate trends that may change after considering incompleteness.\label{fig:result}}
\end{figure}

The number ratio FU/(FU+1O) in each galaxy is directly calculated with the number of Cepheids pulsating in FU and 1O mode in SMC, LMC, the Milky Way, M33 and M31. The number and the ratio of the FU and 1O pulsators in the final Cepheids sample are listed in Table \ref{tab:final}. The metallicity of the galaxies is measured by various works that yield different values. Specifically, the value of $12+\log(O/H)$ is measured to be 8.13 by \citet{1990ApJS...74...93R}, 7.96 by \citet{2015RMxAA..51..135C} and 7.80 by \citet{2019A&A...631A.127M} for SMC, 8.37 by \citet{1990ApJS...74...93R}, 8.33 by \citet{2015RMxAA..51..135C}, 8.50 by \citet{2021ApJ...910...95R} for LMC, 8.71 by \citet{1983MNRAS.204...53S}, 8.70 by \citet{1995RMxAC...3..133E}, 8.66 by \citet{1996ApJ...468L..65S}, 8.73 by \citet{2009ARA&A..47..481A} for the Milky Way, 8.75 by \citet{1997ApJ...489...63G} and 8.76 by \citet{2016MNRAS.458.1866T} for M33, and 9.00 by \citet{1994ApJ...420...87Z} and 8.90 by \citet{2012ApJ...758..133S} for M31. The result is plotted in Figure \ref{fig:result}, where the horizontal line indicates the range of metallicity measured by different works. The result of M33 shows only one point on the figure, because there are only two very consistent measurements. 

The number ratio FU/(FU+1O) is well correlated with the metallicity in that it rises with metallicity, which means that the 1O mode is suppressed in more metal-rich environment. This agrees with the trend of RSGs found by \citet{2019ApJS..241...35R}, and the theoretical prediction by \citet{1998A&A...336..553Y} that the increasing mixing length in metal-rich stars would greatly suppress the growth rate of 1O mode in Cepheids. \citet{2019ApJS..241...35R} calculated the FU/(FU+1O) ratio of RSGs ranges from 0\% to 87\%, while this work yields the ratio within a narrower range from about 50\% to 90\% for Cepheids. This difference may be true for RSGs and Cepheids, while there surely exists some uncertainties. For example, the RSGs sample are incomplete and their mode identification may be uncertain, and the Cepheid samples are also affected by the incompleteness, which will be discussed in the following.

\subsection{Incompleteness of the sample}\label{sec:incompleteness}

For the SMC and LMC sample, the limiting magnitude of the OGLE observation is about 19.5 mag \citep{2018AcA....68..315U}, which is comparable to the faint end of the 1O mode. There could be some incompleteness at the faint end of the sample. The most possibly affected class is the 1O mode of SMC. If this effect is taken into account, the ratio FU/(FU+1O) of SMC would be smaller than the present value, which would then enlarge the difference with LMC and would not change the relativity that the ratio FU/(FU+1O) is higher in the more metal-rich LMC than SMC.

For M31 and M33, the sample of Cepheids cannot be complete. The I-band apparent magnitude distribution of Cepheids in LMC extends to about 18.5 mag at the faint end, which converts to about 24.5 mag in M31 and M33 by taking the difference in distance modulus into account. Meanwhile, the observational limit is about 22 mag (c.f. Figure \ref{fig:division}), i.e $\sim$ two magnitudes brighter than the faint end of possibly both the FU and 1O mode. Again, M33 is more distant and more seriously affected by this incompleteness. If considering the incompleteness, the difference in the ratio FU/(FU+1O) would be larger between M31 and M33 in that it would be decreased more for M33 than for M31. The trend in these two galaxies also agree with the expectation that the 1O mode is suppressed more in a metal-rich galaxy. It may be argued that the luminosity function can be different in M31 and M33 from SMC and LMC, which can alter the completeness. This can be true due to the significant difference in metallicity, but M31 and M33 are similar in galaxy type and metallicity and their luminosity function of Cepheids should be very similar. Though it is unclear whether such incompleteness would lead the ratio FU/(FU+1O) to being smaller than that of the Milky Way, there seems to be no possibility to becoming smaller than the value in SMC and LMC.

For the Milky Way, the sample must be incomplete due to serious extinction in the Galactic plane. The degree of completeness of the FU and 1O mode is hard to judge for the OGLE and ZTF survey.

The above analysis indicates that the trend of the ratio FU/(FU+1O) increasing with metallicity would not be changed with the incompleteness taken into account.

\section{Summary}

The work by \citet{2019ApJS..241...35R} presented the evidence that the pulsation mode of red supergiants depends on the metallicity in that the ratio of fundamental (FU) to first-overtone (1O) mode increases with metallicity. This dependence is explained by the increasing convection in metal-rich atmosphere which suppresses the shallower 1O pulsation. However, red supergiants pulsate generally in semi-regular or even irregular pattern so that the period determination and then the pulsation mode identification suffer relatively large uncertainty.

Cepheids are much more regular variables than red supergiants. The period determination and pulsation mode identification are usually very certain. This work takes Cepheids to examine the dependence of pulsation mode on metallicity. The sample of Cepheids are assembled for SMC, LMC, the Milky Way, M31 and M33 from previous catalogs by the OGLE, DIRECT, PS1, WISE, and ZTF surveys. The division of the FU and 1O mode is refined according to the locations in the P-L diagram of Cepheids. The foreground stars are further examined by the Gaia/EDR3 astrometric information for the extra galaxies, while the contaminants in the Galactic samples are removed by comparing the PLR distance and geometric distance. Based on the final sample, the ratio of FU/(FU+1O) is calculated, being 0.59, 0.60, 0.69, 0.83 and 0.85 for SMC, LMC, the Milky Way, M33 and M31 respectively in order of metallicity. This result proves that the pulsation mode depends on metallicity for Cepheids in the way that the ratio of FU/(FU+1O) increases with metallicity, which is the same as red supergiants. This dependence is not changed if the incompleteness of the samples is taken into account.

\acknowledgments

We would like to thank the anonymous referee for the constructive suggestions that definitely improved this work. We also thank Mr. Tongtian Ren for his help in the double-Gaussian fitting method and Mr. Chuanyu Wei for his help with the Gaia data. This work is supported by the NSFC projects 12133002 and 11533002, National Key R\&D Program of China No. 2019YFA0405503 and CMS-CSST-2021-A09. Z. Z. acknowledges support from National Undergraduate Students Innovation and Entrepreneurship Training Program of China (No. 202110027015). This work has made use of data from the surveys by OGLE, WISE, ZTF, DIRECT, WIYN, Pan-STARRS1, and Gaia.

\software{astropy \citep{2013A&A...558A..33A},
          TOPCAT \citep{2005ASPC..347...29T}}

\bibliography{Cephei_Z}{}
\bibliographystyle{aasjournal}

%

\begin{deluxetable*}{cccccc}[htb]
    \tablecaption{Final sample of Cepheids in each galaxy \label{tab:final}}
    \tablewidth{0pt}
    \tablehead{\colhead{} & \colhead{SMC} & \colhead{LMC}& \colhead{MW} & \colhead{M33}& \colhead{M31}}
    \startdata
    FU & 2667 & 2547 & 1792 & 469 & 1677 \\
    1O & 1878 & 1706 & 821 & 94  & 292\\
    FU/(FU+1O) & 0.59 & 0.60 & 0.69 & 0.83 & 0.85 \\
    \enddata
\end{deluxetable*}

%

\end{CJK*}
\end{document}